# InGaAs/InP single-photon avalanche diodes with 50% detection efficiency and 0.7 kcps dark count rate

DAJIAN CUI,[1,2,3,4,*] WEI CHEN,[2,3,4,5] YONG LEI,[3,4] XUANYI ZHANG,[5,6] QIXIA TONG,[3,4] PENGYU WEN,[3,4] JIANGLIN ZHAO,[3,4,5] XUEFENG YAN,[3,4] YINLIN YAN,[3,4] LI REN,[3,4] CHAO YU,[2,5,6,*] JUN ZHANG,[2,5,6] AND CHUNHUI WANG[1]

[1]*School of Astronautics, Harbin Institute of Technology, Harbin 150001, China*
[2]*Hefei National Laboratory, Hefei 230088, China*
[3]*Chongqing Optoelectronics Research Institute, Chongqing 400060, China*
[4]*Chongqing Key Laboratory of Quantum Information Chips and Devices, Chongqing 400060, China*
[5]*Hefei National Research Center for Physical Sciences at the Microscale and School of Physical Sciences, University of Science and Technology of China, Hefei 230026, China*
[6]*CAS Center for Excellence in Quantum Information and Quantum Physics, University of Science and Technology of China, Hefei 230026, China*
**cuidj@cetccq.com.cn*
**yuch@ustc.edu.cn*

**InGaAs/InP single-photon avalanche diodes (SPADs) are widely used for near-infrared single-photon detection, where high photon-detection efficiency (PDE) and low dark-count rate (DCR) are essential. Here, we design and fabricate InGaAs/InP SPADs with state-of-the-art overall performance. Key improvements include thickening the InGaAs absorption layer, optimizing the guard-ring position, and refining the double-zinc diffusion process. These approaches effectively suppress premature edge breakdown under high electric fields, thereby reducing the DCR while maintaining high PDE. At 233 K under 1.25 GHz gating, the SPAD achieves a PDE of 50.4% at 1550 nm with a DCR of only 0.7 kcps. The proposed device can significantly improve the system performance in near-infrared single-photon detection applications.**

Single-photon detectors (SPDs) are highly sensitive devices for weak-light detection, capable of resolving individual photons—the smallest detectable quanta of electromagnetic radiation [1,2]. Among short-wave infrared detectors, InGaAs/InP single-photon avalanche diodes (SPADs) offer advantages of high performance, low cost, excellent reliability, operational simplicity, compact size, and operation without cryogenic cooling. Consequently, they have been widely used in quantum information and lidar applications [3–6]. These applications generally require a high photon-detection efficiency (PDE) and a low dark-count rate (DCR). In recent years, advances in the understanding of Geiger-mode avalanche mechanisms, together with improvements in device architecture and fabrication technologies, have substantially enhanced the performance of InGaAs/InP SPADs [7–13].

Several schemes have been investigated to improve the performance of InGaAs/InP SPADs. To enhance photon recycling, optical structures such as a Bragg reflector combined with a microlens have been integrated, effectively improving the absorption efficiency [14]. Electric-field engineering has also been exploited to suppress premature edge breakdown. By using selective-area growth to gradually increase the InP layer thickness from the center toward the edge, the edge electric field is reduced and a PDE of 43% at 1550 nm is achieved [15]. Alternatively, thickening the absorption layer can also improve PDE. With thicknesses of 2.0 and 2.2 μm, PDEs of 50% [16] and 55.4% [17] at 1550 nm have been demonstrated, although both devices were operated under low-frequency (1 MHz) gating.

Previously, we fabricated a dielectric–metal reflective layer above the photosensitive region to reflect transmitted photons back into the absorption layer. With a reflectivity of 95%, this structure improved the absorption efficiency by ~20% and resulted in a maximum PDE of 60% at 1550 nm. Under practical operating conditions, the PDE was ~40% with a DCR of 3 kcps [10].

In this work, we report a high-performance InGaAs/InP SPAD with further enhanced PDE and reduced DCR. The thickness of the InGaAs absorption layer is increased from 1.5 to 2.0 μm. Technology computer-aided design (TCAD) simulations are used to optimize the double-zinc-diffusion process and guard-ring position. The fabricated SPAD is packaged in a butterfly housing with integrated thermoelectric cooling. Characterization results show that the device achieves a PDE of 50.4% and a DCR of 0.7 kcps at an operating temperature of 233 K.

The InGaAs/InP SPAD employs a separated absorption, grading, charge, and multiplication (SAGCM) heterostructure based on our previous design [18], with the InGaAs absorption-layer thickness increased to 2.0 μm. To suppress premature edge breakdown and reduce the dark count rate while maintaining a high PDE, TCAD simulations are performed to optimize the junction shape, including the depth of the double-zinc-diffused junction (Δy) and the lateral distance between the guard ring and the junction (Δx).

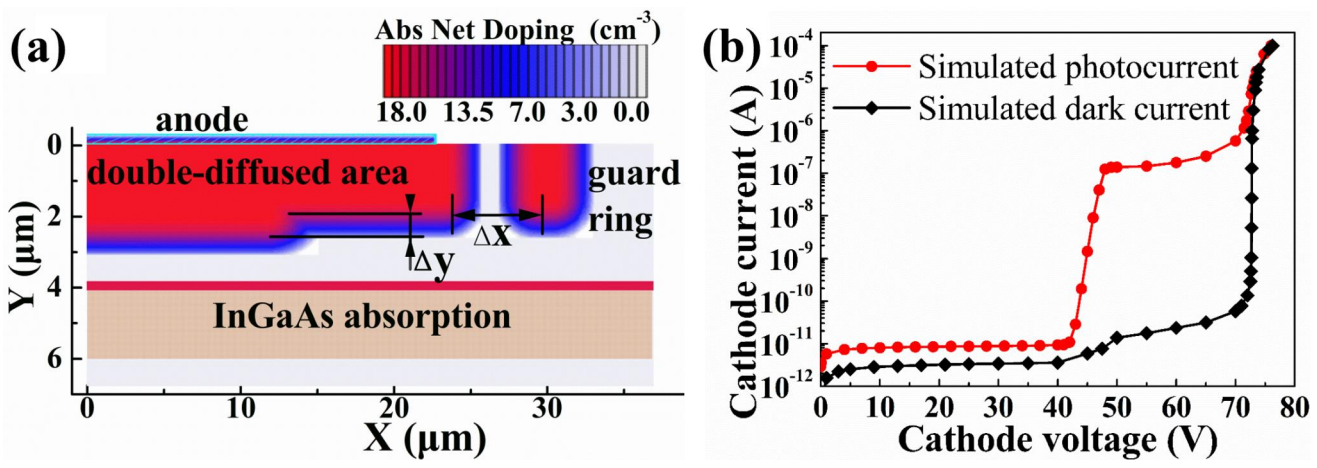

Fig. 1. (a) Simulated net doping distribution of the InGaAs/InP SPAD with a 25 μm active-area diameter, showing the double-diffused junction, the guard ring, the deep–shallow junction depth difference Δy, and the guard-ring-to-junction spacing Δx; (b) Simulated photocurrent under 1550-nm illumination and dark current as functions of the cathode voltage.

Fig. 1(a) presents the simulated structure of the InGaAs/InP SPAD with a 25 μm active area diameter. The deep and shallow junction profiles, along with the doping concentration distributions used to model the double-zinc diffusion process and the guard ring, are clearly indicated in the figure. The simulated diffused junction exhibits a stepped doping concentration profile, with a maximum doping concentration of ~$1 \times 10^{18}$ $cm^{-3}$ at the InP top surface. Fig. 1(b) presents the simulated photocurrent and dark current characteristics under 1550 nm illumination as a function of the applied bias voltage. The simulated device exhibits a punch-through voltage ($V_{pt}$) of ~42 V, a breakdown voltage ($V_{br}$) of ~72 V, and a dark current of ~0.2 nA at 71 V and 300 K.

In planar double-diffused InGaAs/InP SPADs, the depth difference, Δy, between the deep and shallow junctions is a critical design parameter. Reducing Δy lowers the electric field in the multiplication region, thereby suppressing trap-assisted tunneling and reducing the DCR. A smaller Δy also mitigates charge persistence [19]. However, if Δy is too small, the electric field beneath the shallow diffusion becomes excessively strong, leading to premature breakdown and an increase in the DCR [20]. Therefore, the design objective is to minimize Δy while avoiding premature breakdown.

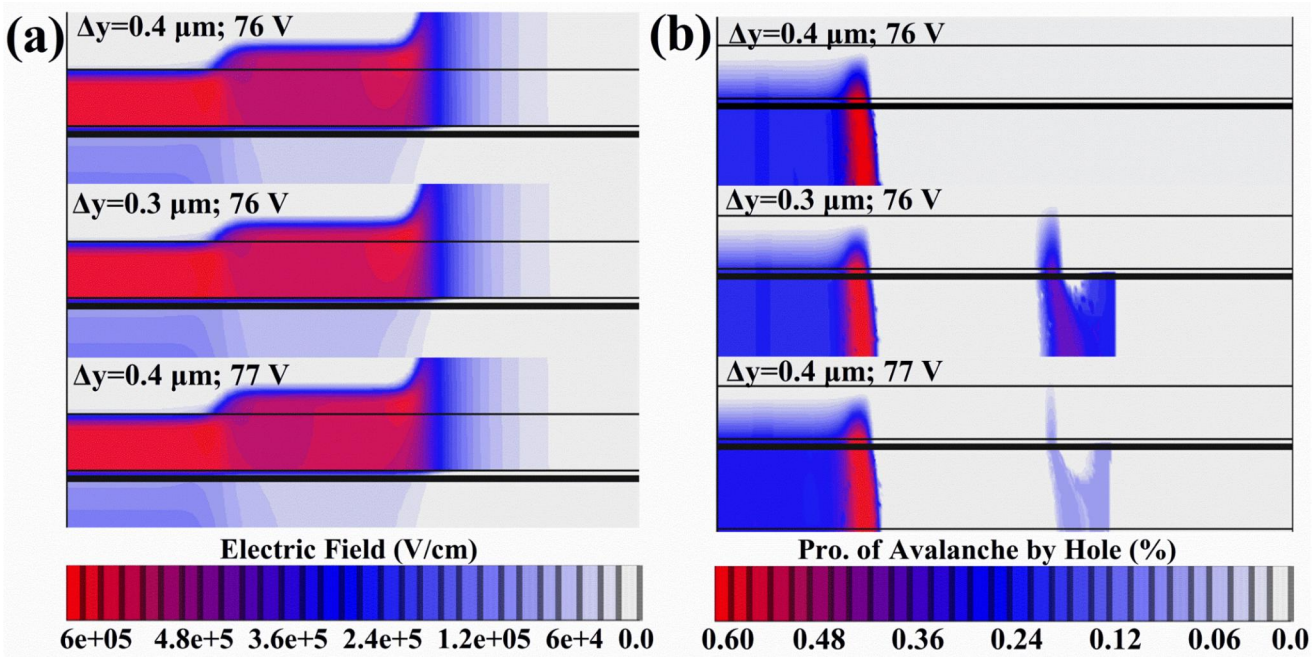

Fig. 2. Simulated two-dimensional (a) electric-field intensity and (b) avalanche-probability distributions for various deep–shallow junction depth differences.

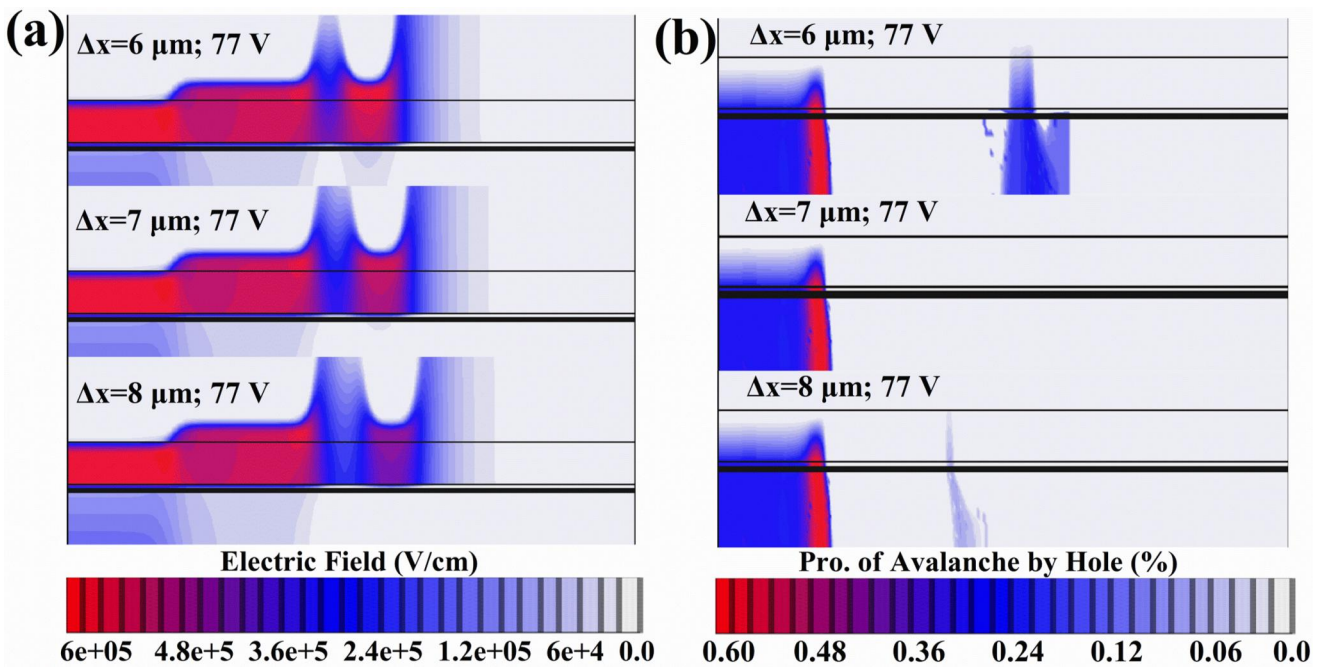

Fig. 3. Simulated two-dimensional (a) electric-field intensity and (b) avalanche-probability distributions for different guard-ring-to-junction spacings.

At a bias voltage of 76 V, the electric-field intensity and avalanche-probability distributions are simulated as Δy is reduced step by step. As shown in Fig. 2, Δy must be at least 0.4 μm to prevent premature breakdown. When Δy is further reduced to 0.3 μm, premature breakdown occurs near the shallow junction. Accordingly, Δy is set to 0.4 μm in the device design. Nevertheless, premature breakdown still occurs when the bias voltage is increased to 77 V, even with this optimized Δy.

To further reduce the DCR at high bias voltages, a floating guard ring is introduced to share the peripheral voltage drop. Its field-relief effect depends critically on the spacing, Δx, between the guard ring and the main junction. If Δx is too small, the electric field crowds into the narrow gap between the guard ring and the main junction. Conversely, if Δx is too large, the guard ring becomes weakly coupled to the main junction and can no longer effectively share the voltage drop. An optimal spacing therefore exists. To determine this spacing, the edge electric-field and avalanche-probability distributions are simulated for different values of Δx.

Fig. 3 presents the simulated electric-field intensity and avalanche-probability distributions at a bias voltage of 77 V for different guard-ring-to-junction spacings. As Δx increases from 6 to 7 μm, the avalanche probability near the shallow junction progressively decreases to zero, indicating that a spacing of 7 μm effectively suppresses premature breakdown in this region. Further increasing Δx to 8 μm causes avalanche events to reappear, with the avalanche probability approaching that of a device without a guard ring. Therefore, Δx = 7 μm is selected for the device design.

Based on the optimized design parameters, the InGaAs/InP SPAD is fabricated using a planar process based on the SAGCM epitaxial structure with a 2.0 μm-thick InGaAs absorption layer. Fig. 4(a) shows a micrograph of the fabricated chip (frontside). The chip is then mounted in a compact butterfly package with an integrated thermoelectric cooler (TEC), as shown in Fig. 4(b), which can stabilize the chip temperature down to 233 K. The package is pigtailed with a single-mode fiber terminated with an FC/PC connector for convenient coupling to the 1550 nm optical test system.

Then, the current–voltage (I–V) characteristics of the packaged SPAD are measured under 0.1 μW of 1550 nm continuous illumination and in the dark at 300 K and 233 K, as shown in Fig. 4(c). At 300 K, the device exhibits a punch-through voltage ($V_{pt}$) of 42 V and a breakdown voltage ($V_{br}$) of 72 V. Both values agree well with the simulated results in Fig. 1(b), confirming that the doping profile and junction geometry of the fabricated device closely reproduce the optimized design. As the temperature decreases from 300 K to 233 K, $V_{br}$ shifts from 72 V to 65 V. At a reverse bias of $V_{br} - 1$ V, the dark current is as low as ~20 pA at 233 K. The well-behaved I–V characteristics demonstrate the high quality of the double-diffused junction and the effectiveness of the guard-ring design.

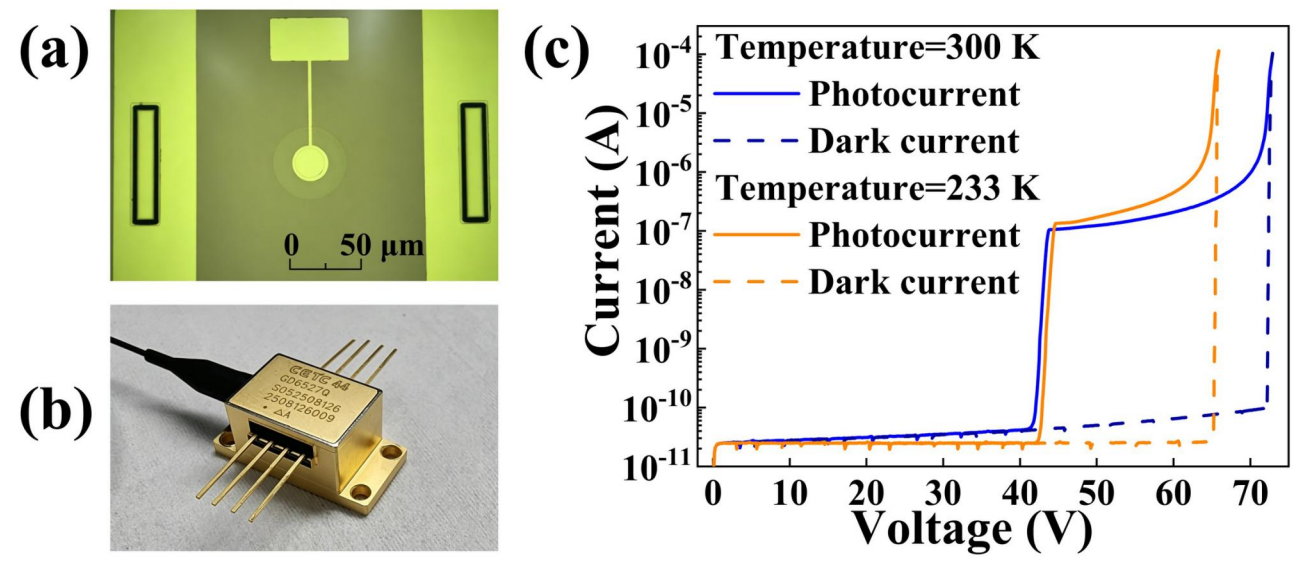

Fig. 4. (a) Front-side micrograph of the fabricated InGaAs/InP SPAD chip. (b) Photograph of the packaged SPAD with integrated cooling. (c) Measured current–voltage characteristics at 300 and 233 K under 0.1 µW continuous-wave illumination at 1550 nm and in the dark.

The Geiger-mode performance of the InGaAs/InP SPAD is characterized using a 1.25 GHz sine-wave gating system, as shown in Fig. 5(a). A 1550 nm pulsed laser operating at a repetition rate of 625 kHz and a pulse width of 20 ps is attenuated by a variable optical attenuator (VOA) to an average of 0.1 photon per pulse before entering the SPAD. Meanwhile, the laser provides a synchronous clock to a digital signal generator (DSG), which produces a 1.25 GHz sine wave. The sine wave is then amplified by a power amplifier (PA) and applied to the SPAD as the gating signal. The avalanche response of the SPAD is read out by a monolithically integrated readout circuit (MIRC) [13], which performs low-noise amplification and filtering on the avalanche signal, and the resulting counts are recorded by a time-correlated single-photon counting (TCSPC) system. Throughout the measurement, the DC bias and the device temperature are maintained by a source meter and a TEC controller, respectively.

In sine-wave gating mode, the excess bias voltage (Vex) is defined as $\text{Vex} = V_{dc} + V_{gate} - V_{br}$, where $V_{dc}$ is the DC bias voltage, and $V_{gate}$ is the amplitude of the sine-wave gate, which is fixed at 6 V throughout the experiments. Fig. 5(b) shows the DCR as a function of PDE, with the values of Vex labeled beside each data point. As Vex increases, the DCR increases with PDE exponentially. Benefiting from the optimized double-diffusion profile and the guard-ring design, the device maintains an exceptionally low DCR even at high detection efficiency. With a Vex of 5.5 V and a temperature of 233 K, the PDE reaches a maximum of 50.4% at 1550 nm, while the DCR is only 0.7 kcps. The afterpulse probability (Pap) is extracted from the TCSPC histogram with a hold-off time of 100 ns. Pap is calculated as $\text{Pap}=(C_{total}-C_{photon}-C_{dark})/C_{laser}$, where $C_{total}$ represents the total counts, $C_{photon}$ represents the photon counts obtained from the main peak of the TCSPC histogram, and $C_{dark}$ represents the dark counts. Fig. 5(c) shows the Pap as a function of PDE. At the operating point of 233 K and 50% PDE, the device exhibits a Pap of ~8.4%. Fig. 5(d) shows the typical TCSPC histogram over a time range of 0–400 ns, with timing jitter of the SPD is shown in the inset. In sine-wave gating mode, the SPD produces no output during gate-off periods, so the timing distribution is divided into discrete peaks corresponding to individual gates. With 50% PDE, the system timing jitter, including the laser pulse, characterization system, and SPAD device, is ~150 ps full width at half maximum (FWHM), which is well suited for quantum communication and lidar applications.

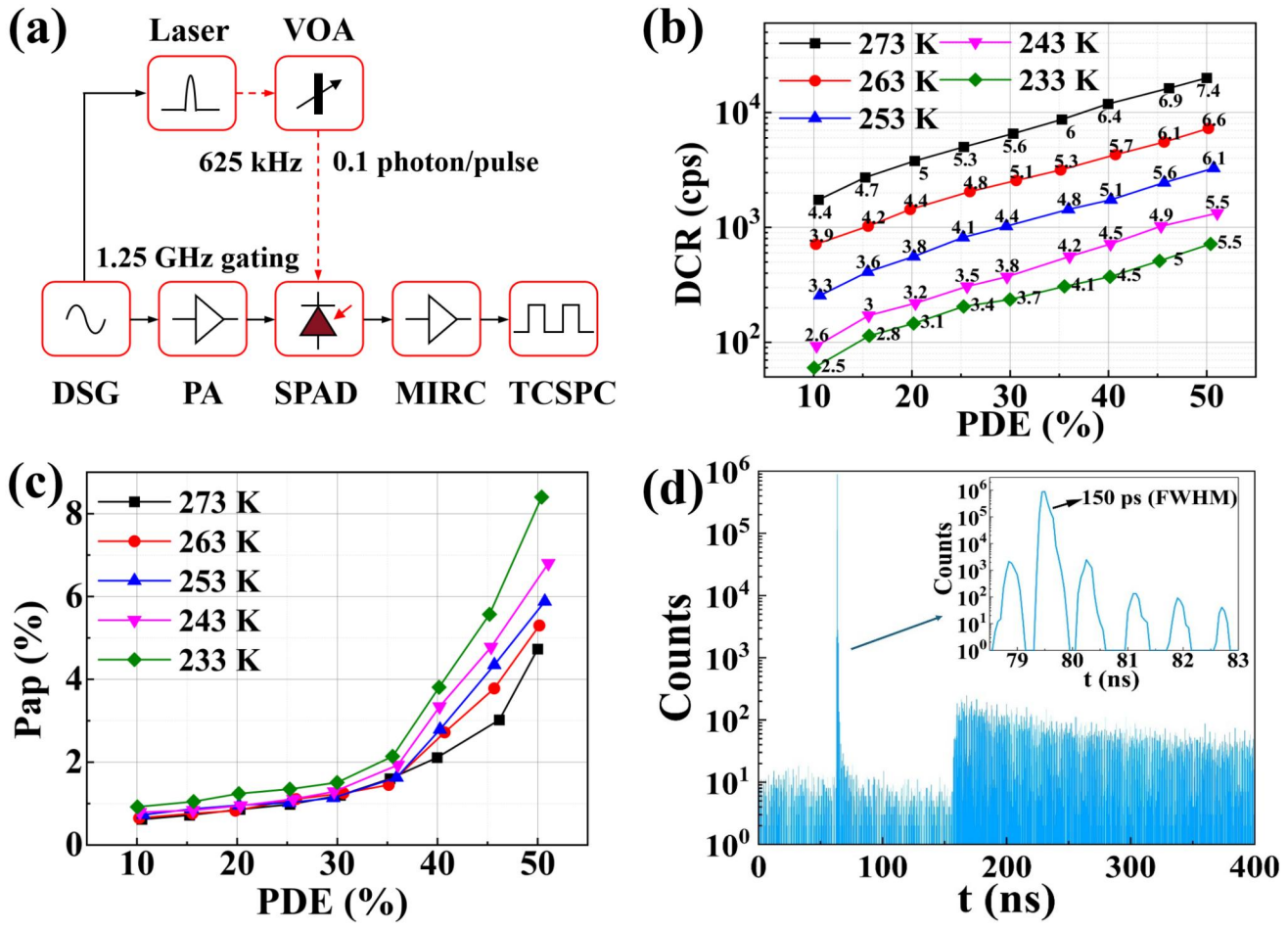

Fig. 5. (a) Schematic of the 1.25-GHz sine-wave-gating characterization system (VOA: variable optical attenuator; DSG: digital signal generator; PA: power amplifier; MIRC: monolithically integrated readout circuit; TCSPC: time-correlated single-photon counting). (b) DCR versus PDE at different temperatures. The values of Vex are labeled beside each data point in volts. (c) Afterpulse probability versus PDE at different temperatures with a 100-ns hold-off time. (d) Temporal response of the SPAD measured with a pulsed laser at 50% PDE.

**Table 1. COMPARISON OF INGAAS/INP SPADS**

| Ref. | Gating Frequency | T (K) | PDE@1550 nm | DCR (kcps) | DCR/gate |
|---|---|---|---|---|---|
| [10] | 1.25 GHz | 300 | 60% | 340 | $2.7\times10^{-4}$ |
| [16] | 1 MHz | 225 | 50% | 20 | $2.0\times10^{-2}$ |
| [21] | 1.27 GHz | 289 | 55.9% | 597 | $4.7\times10^{-4}$ |
| [22] | 10 MHz | 293 | 21% | 0.1 | $1.0\times10^{-5}$ |
| [23] | 1.25 GHz | 243 | 30% | 1.6 | $1.3\times10^{-6}$ |
| This work | 1.25 GHz | 233 | 50.4% | 0.7 | $5.6\times10^{-7}$ |

In summary, we demonstrated a high-performance InGaAs/InP SPAD featuring a 2.0-µm-thick absorption layer, a TCAD-optimized double-zinc-diffusion profile, and a properly positioned floating guard ring. This design effectively suppresses premature edge breakdown, enabling low-noise operation at high excess bias. Under 1.25-GHz sine wave gating at 233 K, the SPAD achieves a PDE of 50.4% at 1550 nm, a DCR of 0.7 kcps, an afterpulse probability of 8.4%, and a timing jitter of 150 ps. As summarized in Table 1, it offers an excellent trade-off between detection efficiency and dark count rate among reported InGaAs/InP SPADs, making the device a competitive candidate for practical near-infrared single-photon detection.

**Funding.** This work is supported by the Quantum Science and Technology-National Science and Technology Major Project under Grant 2021ZD0300803.

**Disclosures.** The authors declare no conflicts of interest.

**Data availability.** Data underlying the results presented in this paper are not publicly available at this time but may be obtained from the authors upon reasonable request.

## References

[1] R. H. Hadfield, Nature Photon. 3, 696–705 (2009).

[2] M. D. Eisaman, J. Fan, A. Migdall, and S. V. Polyakov, Rev. Sci. Instrum. 82, 071101 (2011).

[3] F. Acerbi, M. Anti, A. Tosi, and F. Zappa, IEEE Photonics J. 5, 6800209 (2013).

[4] J. Zhang, M. A. Itzler, H. Zbinden, and J.-W. Pan, Light Sci. Appl. 4, e286 (2015).

[5] C. Yu, Q. Xu, and J. Zhang, Meas. Sci. Technol. 35, 122003 (2024).

[6] M.-J. Shangguan, J.-W. Qiu, J.-L. Yuan, Z.-F. Shu, L.-F. Zhou, and H.-Y. Xia, Front. Remote Sens. 2, 787111 (2022).

[7] X.-L. Liang, J.-H. Liu, Q. Wang, D.-B. Du, J. Ma, G. Jin, Z.-B. Chen, J. Zhang, and J.-W. Pan, Rev. Sci. Instrum. 83, 083111 (2012).

[8] A. Bouzid, A. M. Nahhas, and K. Guedri, Opt. Commun. 328, 37-40 (2014).

[9] C. Scarcella, G. Boso, A. Ruggeri, and A. Tosi, IEEE J. Sel. Top. Quantum Electron. 21, 3800306 (2015).

[10] Y.-Q. Fang, W. Chen, T.-H. Ao, C. Liu, L. Wang, X.-J. Gao, J. Zhang, and J.-W. Pan, Rev. Sci. Instrum. 91, 083102 (2020).

[11] W.-H. Jiang, X.-J. Gao, Y.-Q. Fang, J.-H. Liu, Y. Zhou, L.-Q. Jiang, W. Chen, G. Jin, J. Zhang, and J.-W. Pan, Rev. Sci. Instrum. 89, 123104 (2018).

[12] Q. Xu, C. Yu, D. Cui, X.-Y. Zhang, W. Chen, Y.-Q. Fang, L. Jiang, Q. Tong, J. Zhao, and J. Zhang, IEEE J. Sel. Top. Quantum Electron. 31, 3801007 (2025).

[13] Q. Xu, C. Yu, W. Chen, J. Zhao, D. Cui, J. Zhang, and J.-W. Pan, IEEE J. Sel. Top. Quantum Electron. 30, 6400107 (2024).

[14] B.-J. Zhang, S.-Z. Yin, Y.-J. Liu, Z.-J. Jiang, W. He, Q.-W. Li, J.-W. Hao, and L. Wang, J. Lightwave Technol. 40, 3832-3838 (2022).

[15] E. Kizilkan, U. Karaca, V. Pesic, M.-J. Lee, C. Bruschini, A. J. SpringThorpe, A. W. Walker, C. Flueraru, O. J. Pitts, and E. Charbon, IEEE J. Sel. Top. Quantum Electron. 28, 9300209 (2022).

[16] F. Signorelli, F. Telesca, E. Conca, A. Della Frera, A. Ruggeri, A. Giudice, and A. Tosi, in Proc. IEEE Int. Electron Devices Meet. (IEDM) 20.3.1 (2021).

[17] T.-T. He, X.-H. Yang, Y.-S. Tang, R. Wang, and Y.-J. Liu, J. Semicond. 43, 102301 (2022).

[18] J. Ma, B. Bai, L.-J. Wang, C.-Z. Tong, G. Jin, J. Zhang, and J.-W. Pan, Appl. Opt. 55, 7497-7502 (2016).

[19] F. Telesca, F. Signorelli, and A. Tosi, IEEE J. Sel. Top. Quantum Electron. 30, 3801207 (2024).

[20] F. Signorelli, F. Telesca, E. Conca, A. Della Frera, A. Ruggeri, A. Giudice, and A. Tosi, IEEE J. Sel. Top. Quantum Electron. 28, 3801310 (2022).

[21] A. Tada, N. Namekata, and S. Inoue, Jpn. J. Appl. Phys. 59, 072004 (2020).

[22] S.-H. Baek, S.-C. Yang, C.-Y. Park, C.-W. Park, S.-B. Cho, and S.-W. Ryu, J. Korean Phys. Soc. 78, 634-641 (2021).

[23] Y.-B. Fan, T.-T. Shi, W.-J. Ji, L. Zhou, Y. Ji, and Z.-L. Yuan, Opt. Express 31, 7515-7522 (2023).